# Clique Matrices for Statistical Graph Decomposition and Parameterising Restricted Positive Definite Matrices

**David Barber**
Department of Computer Science, University College London
www.cs.ucl.ac.uk/staff/d.barber

## Abstract

We introduce Clique Matrices as an alternative representation of undirected graphs, being a generalisation of the incidence matrix representation. Here we use clique matrices to decompose a graph into a set of possibly overlapping clusters, defined as well-connected subsets of vertices. The decomposition is based on a statistical description which encourages clusters to be well connected and few in number. Inference is carried out using a variational approximation. Clique matrices also play a natural role in parameterising positive definite matrices under zero constraints on elements of the matrix. We show that clique matrices can parameterise all positive definite matrices restricted according to a decomposable graph and form a structured Factor Analysis approximation in the non-decomposable case.

## 1 Introduction

Undirected graphs may be used to represent connectivity or adjacency structures in data. For example, in Collaborative Filtering, the nodes(vertices) in the graph may represent products, and a link(edge) between nodes $i$ and $j$ could be used to indicate that customers who by product $i$ frequently also buy product $j$. This paper concerns decomposing the graph into well-connected clusters of nodes[1]. In Fig.1a product 3 is typically bought along with products 1 and 2, or with products 4 and 5, though these two product-groups are otherwise disjoint. A formal specification of the problem of finding a minimum number of well-connected subsets is to phrase this as MIN CLIQUE COVER[9, 17]. However, in some applications, provided that only a small number of links in an 'almost clique' are missing, this may be considered a sufficiently well-connected group of nodes to form a cluster. We will therefore develop a statistical technique to reveal clusters of nodes and to identify the smallest number of such clusters.

---
[1]Not to be confused with graph-partitioning.

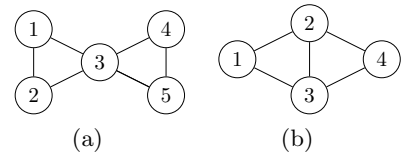

Figure 1: Two simple undirected graphs

Our main contribution is the introduction of the *clique matrix* formalism, a generalisation of the incidence matrix. We apply this to the clustering problem, in addition to demonstrating an application in constrained covariance parameterisation.

## 2 Clique Decomposition

The symmetric adjacency matrix has elements $A_{ij} \in \{0,1\}$, with a 1 indicating a link between nodes $i$ and $j$. For the graph in Fig.1b, the adjacency matrix is

$$A = \begin{pmatrix} 1 & 1 & 1 & 0 \\ 1 & 1 & 1 & 1 \\ 1 & 1 & 1 & 1 \\ 0 & 1 & 1 & 1 \end{pmatrix} \qquad (1)$$

where we include self connections on the diagonal. Given a graph $G$ with adjacency matrix $A$, our aim is to find a 'simpler' description of $A$ that reveals underlying cluster structure.

### 2.1 Two-Clique Decomposition

Given the undirected graph in Fig.1b, the *incidence matrix* $Z_{inc}$ is an alternative description of the adjacency structure[6]. Given the $V$ nodes in the graph, we construct $Z_{inc}$ as follows: For each link $ij$ in the graph, form a column of the matrix $Z_{inc}$ with zero entries except for a 1 in the $i^{th}$ and $j^{th}$ row. The column ordering is arbitrary. For example, on the left

$$Z_{inc} = \begin{pmatrix} 1 & 1 & 0 & 0 & 0 \\ 1 & 0 & 1 & 1 & 0 \\ 0 & 1 & 1 & 0 & 1 \\ 0 & 0 & 0 & 1 & 1 \end{pmatrix}, Z_{inc} Z_{inc}^\mathsf{T} = \begin{pmatrix} 2 & 1 & 1 & 0 \\ 1 & 3 & 1 & 1 \\ 1 & 1 & 3 & 1 \\ 0 & 1 & 1 & 2 \end{pmatrix}$$

is an incidence matrix for the graph in Fig.1b. Taking the outer-product with itself, on the right, we see that[2]

$$A = H\left(Z_{inc} Z_{inc}^\mathsf{T}\right) \qquad (2)$$

where $[H(M)]_{ij} = 1$ if $M_{ij} > 0$ and is 0 otherwise (i.e. $H(\cdot)$ is the element-wise Heaviside step function).

A useful viewpoint of the incidence matrix is that it identifies two-cliques in the graph (here we are using the term 'clique' in the non-maximal sense). There are five 2-cliques in Fig.1b, and each column of $Z_{inc}$ specifies which elements are in each 2-clique.

### 2.2 Clique matrices

The incidence matrix can be generalised to describe larger cliques. Consider the following matrix as a decomposition for Fig.1b, and its outer-product:

$$Z = \begin{pmatrix} 1 & 0 \\ 1 & 1 \\ 1 & 1 \\ 0 & 1 \end{pmatrix}, \qquad ZZ^\mathsf{T} = \begin{pmatrix} 1 & 1 & 1 & 0 \\ 1 & 2 & 2 & 1 \\ 1 & 2 & 2 & 1 \\ 0 & 1 & 1 & 1 \end{pmatrix} \qquad (3)$$

The interpretation is that $Z$ represents a decomposition into two 3-cliques. As in the incidence matrix, each column represents a clique, and the rows containing a '1' express which elements are in the clique defined by that column. Both $Z_{inc}$ and $Z$ satisfy

$$A = H\left(ZZ^\mathsf{T}\right) = H\left(Z_{inc} Z_{inc}^\mathsf{T}\right) \qquad (4)$$

for Fig.1b. For clustering, $Z$ is to be preferred against the incidence decomposition, since $Z$ decomposes the graph into a smaller number of larger cliques. Indeed, $Z$ solves MIN CLIQUE COVER for Fig.1b.

**Definition 1** (Clique Matrix). *Given an adjacency matrix $[A]_{ij}, i, j = 1, \ldots, V$ ($A_{ii} = 1$), a clique matrix $Z$ has elements $Z_{i,c} \in \{0, 1\}, i = 1, \ldots, V, c = 1, \ldots C$ such that $A = H(ZZ^\mathsf{T})$.*

A Clique Matrix $Z \in \{0,1\}^{V \times C}$ is *minimal* for $A$ if there exists no other clique matrix for $Z \in \{0,1\}^{V \times C'}$ with a smaller number of columns $C' < C$.

That each column of $Z$ expresses a clique is clear from

$$\left[ZZ^\mathsf{T}\right]_{ij} = \sum_k Z_{ik} Z_{jk} \qquad (5)$$

---
[2]$(\cdot)^\mathsf{T}$ represents matrix transpose.

For each $k$, the nodes $i$ and $j$ corresponding to 1's in the $k^{th}$ column of $Z$ give a product $Z_{ik}Z_{jk} = 1$. Since this happens for every non-zero $i$ and $j$ pair in the $k^{th}$ column, all of the pair connections give rise to a non-zero product. In other words, all the nodes corresponding to non-zero elements of the $k^{th}$ column are connected to each other, thus forming a clique.

The interpretation of the elements of $ZZ^\mathsf{T}$ is that diagonal elements $\left[ZZ^\mathsf{T}\right]_{ii}$ express the number of cliques/columns that vertex $i$ occurs in. Off-diagonal elements $\left[ZZ^\mathsf{T}\right]_{ij}$ contain the number of cliques/columns that vertices $i$ and $j$ jointly inhabit.

Whilst finding a clique decomposition $Z$ is easy (use the incidence matrix for example), finding a clique decomposition with the minimal number of columns, i.e. solving MIN CLIQUE COVER, is NP-Hard[9]. One approach would be to use a recursive procedure that searches for maximal cliques in the graph or related techniques based on finding large densely connected subgraphs[17]. The route that we take here is different and motivated by the idea that perfect clique decomposition is not necessarily desirable if the aim is only to find well-connected clusters in $G$.

## 3 Statistical Clique Decompositions

To find 'well-connected' clusters, we relax the constraint that the decomposition is in the form of cliques in the original graph. Our approach is to view the absence of links as statistical fluctuations away from a perfect clique.

Given a $V \times C$ matrix $Z$, we desire that the higher the overlap between rows[3] $z_i$ and $z_j$ is, the greater the probability of a link between $i$ and $j$. This may be achieved using, for example,

$$p(i \sim j | Z) = \sigma\left(z_i z_j^\mathsf{T}\right) \qquad (6)$$

where we define

$$\sigma(x) \equiv \left(1 + e^{\beta(0.5 - x)}\right)^{-1} \qquad (7)$$

and $\beta$ controls the steepness of the function. The 0.5 shift in Eq. (7) ensures that $\sigma$ approximates the step-function, since the argument of $\sigma$ is an integer. Under Eq. (6), if $z_i$ and $z_j$ have at least one '1' in the same position, $z_i z_j^\mathsf{T} - 0.5 > 0$ and $p(i \sim j)$ is high. Absent links contribute $p(i \not\sim j | Z) = 1 - p(i \sim j | Z)$. $\beta$ controls how strictly $\sigma(ZZ^\mathsf{T})$ matches $A$; for large $\beta$, very little flexibility is allowed and only cliques will be identified. For small $\beta$, subsets that would be cliques if it were not for a small number of missing links, are clustered together. The setting of $\beta$ is user and problem dependent.

---
[3]We use lower indices $z_i$ to denote the the $i^{th}$ row of $Z$.

Given $Z$, and assuming each element of the adjacency matrix is sampled independently from the generating process, the joint probability of observing $A$ is (neglecting its diagonal elements),

$$p(A|Z) = \prod_{i \sim j} \sigma\left(z_i z_j^\mathsf{T}\right) \prod_{i \nsim j} \left(1 - \sigma\left(z_i z_j^\mathsf{T}\right)\right)$$

The ultimate quantity of interest is the posterior,

$$p(Z|A) \propto p(A|Z) p(Z) \qquad (8)$$

where $p(Z)$ is a prior over clique matrices. Later we place a prior on $Z$ to encourage the smallest number of clusters to be identified (and hence for the size of the clusters to be large). However, since finding such $Z$, even in the case of a fixed desired number of clusters, $C$, is hard, we develop an algorithm to approximately discover clique matrices, before discussing non-uniform priors $p(Z)$.

## 4 Finding $Z$ for a fixed cluster number

Formally, our task is to find the Most likely A Posteriori (MAP) solution $\arg\max_Z p(A|Z)$ where $Z$ is a $V \times C$ binary matrix. A variety of deterministic and randomised methods could be brought to bear on this problem. The approach we take here is to approximate the marginal posterior $p(z_{ij}|A)$ and then to assign each $z_{ij}$ to that state which maximises this posterior marginal (MPM). This has the advantage of being closely related to marginal likelihood computations, which will prove useful later for addressing the issue of finding the number of clusters. Here we develop a straightforward variational approach based on a simple factorised approximation to the posterior.

### 4.1 Mean Field Approximation

Given the intractable $p(Z|A) \propto p(A|Z)$, a fully factorised mean-field approximation (see, e.g. [20])

$$q(Z) = \prod_{i=1}^V \prod_{c=1}^C q(z_{i,c}) \qquad (9)$$

can be found by minimising the KL divergence

$$KL(q,p) = \langle \log q \rangle_q - \langle \log p \rangle_q \qquad (10)$$

where $\langle \cdot \rangle_q$ represents expectation with respect to $q$. The first 'entropic' term simply decomposes into $\sum_{i,c} \langle \log q(z_{i,c}) \rangle$. The second, 'energy' term, up to a constant is

$$\sum_{i \sim j} \left\langle \log \sigma \left( \sum_c z_{ic} z_{j,c} \right) \right\rangle_q$$
$$+ \sum_{i \nsim j} \left\langle \log \left( 1 - \sigma \left( \sum_c z_{ic} z_{j,c} \right) \right) \right\rangle_q \qquad (11)$$

The first term of Eq. (11) encourages graph links to be preserved under the decomposition, and is given by

$$\sum_{i \sim j} \left\langle f\left( \sum_{d=1}^C z_{id} z_{jd} \right) \right\rangle_{\prod_{e=1}^C q(z_{ie}) q(z_{je})} \qquad (12)$$

where $f(x) \equiv \log \sigma(x)$. Minimising Eq. (10) can be achieved by differentiation. Differentiating the energy contribution from the present links, Eq. (12) with respect to $q(z_{kc})$ we identify two cases: when $i=k$ and when $j=k$. Due to symmetry, the derivative is

$$2 \sum_{k \sim j} \left\langle f\left( \sum_d z_{kd} z_{jd} \right) \right\rangle_{\prod_e q(z_{je}) \prod_{g \neq c} q(z_{kg})} \equiv \Psi(Q) \qquad (13)$$

Similarly, the derivative of the absent-links energy is

$$2 \sum_{k \nsim j} \left\langle f'\left( \sum_d z_{kd} z_{jd} \right) \right\rangle_{\prod_e q(z_{je}) \prod_{g \neq c} q(z_{kg})} \equiv \Psi'(Q) \qquad (14)$$

where $f'(x) \equiv \log(1 - \sigma(x))$. Equating the derivative of Eq. (10) to zero, a fixed point condition for each $q_{k,c}$ $k=1,\ldots,V, c=1,\ldots,C$ is

$$q(z_{kc}) \propto e^{\Psi(Q) + \Psi'(Q)} \qquad (15)$$

A difficulty here is that neither $\Psi(Q)$ nor $\Psi'(Q)$ are easy to compute, due to the non-linearities. A simple Gaussian Field approximation[3] assumes $\sum_d z_{kd} z_{jd}$ is Gaussian distributed for a fixed state of $z_{i,c}$. In this case, we need to find the mean and variance of $\sum_d z_{kd} z_{jd}$. Writing $\theta_{ab} \equiv q(z_{ab} = 1)$, and using the independence of $q$, the mean is given by

$$\mu_{kj} = z_{kc} \theta_{jc} + \sum_{d \neq c} \theta_{kd} \theta_{jd}$$

A similar expression is easily obtained for the variance $\sigma_{kj}^2$. The Gaussian Field approximation then becomes,

$$q(z_{kc}) \propto e^{2 \langle \sum_{j \sim k} f(x) + \sum_{j \nsim k} f'(x) \rangle_{\mathcal{N}(x|\mu_{kj}, \sigma_{kj}^2)}} \qquad (16)$$

where the one dimensional averages are performed numerically. By evaluating Eq. (16) for the two states of $z_{kc}$ (and noting that the mean and variance of the field depends on these states), the approximate update for $\theta_{kc}$ is obtained. A simpler alternative is to assume that the variance of the field is zero, and approximate the averages by evaluating the functions at the mean of the field. We found that this latter procedure often gives satisfactory performance and therefore used this simpler and faster approach in the experiments.

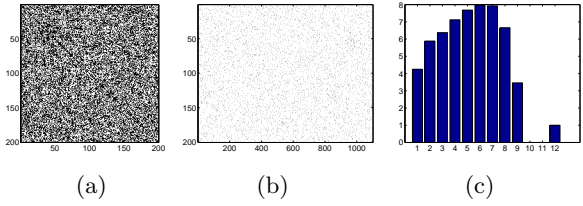
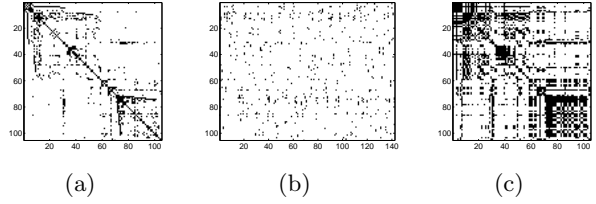

Figure 2: (a) Adjacency matrix for the DIMACS `brock200-2` `MAX-CLIQUE` challenge. Black denotes the presence of a link. (b) Clique Matrix. (c) Log$_2$-histogram of clique occurrence (+1); correctly solves `MAX-CLIQUE` (12) as well as identifying all remaining clusters.

Figure 3: (a) Adjacency matrix of 105 Political Books (black=1). (b) Clique matrix: 521 non-zero entries. (c) Adjacency reconstruction using an Approximate Clique Matrix with 10 cliques – see also Fig. 4.

One epoch corresponds to updating all the $\theta_{kc}, k = 1, \ldots, V, c = 1 \ldots, C$. During each epoch the order in which the parameters are updated is chosen randomly.

## 5 Finding the number of clusters

To bias the contributions to $A$ to occur from a small number of columns of $Z$, we first reparameterize $Z$ as

$$Z = \left( \alpha_1 z^1, \ldots, \alpha_{C_{max}} z^{C_{max}} \right) \quad (17)$$

where $\alpha_c \in \{0, 1\}$ play the role of indicators and $z^c$ is the vector of column $c$ of $Z$. $C_{max}$ is an assumed maximal number of clusters we desire to find. Ideally, we would like to find a likely solution $Z$ with a low number of indicators $\alpha_1, \ldots, \alpha_{C_{max}}$ in state 1. To achieve this we define a prior on $\alpha$[4],

$$p(\alpha|\nu) = \prod_c \nu^{I[\alpha_c=1]} (1-\nu)^{I[\alpha_c=0]} \quad (18)$$

To encourage a small number of $\alpha'$s to be used, we use a Beta prior $p(\nu)$. This gives rise to a Beta-Bernoulli distribution

$$p(\alpha) = \int_\nu p(\alpha|\nu)p(\nu) = \frac{B(a+N, b+C_{max}-N)}{B(a,b)} \quad (19)$$

where $B(a, b)$ is the normalisation constant of the beta-distribution. $N \equiv \sum_{c=1}^{C_{max}} I[\alpha_c = 1]$, namely the number of indicators in state 1. To encourage strongly that a small number of components should be active, we set $a = 1, b = 3$. Through Eq. (17), the prior on $\alpha$ thus induces a prior on $Z$. The resulting distribution $p(Z, \alpha|A) \propto p(Z|\alpha)p(\alpha)$ is formally intractable.

---
[4]$I[x=y]$ is 1 if $x = y$ and 0 otherwise.

### 5.1 Variational Bayes

To deal with the intractable joint posterior we adopt a similar strategy to the fixed $C$ case and employ a variational procedure to seek a factorised approximation $p(\alpha, Z|A) \approx q(\alpha)q(Z)$ based on minimising

$$KL(q(\alpha)q(Z), p(\alpha, Z|A)) \quad (20)$$

$q(Z)$ **updates**

A fixed point condition for the optimum of Eq. (20) is

$$q(Z) \propto e^{\langle \log p(A|Z,\alpha) \rangle_{q(\alpha)}} \approx e^{\log p(A|Z, \langle \alpha \rangle)} \quad (21)$$

The average over $q(\alpha)$ in Eq. (21) in the first expression is complex to carry out and we simply approximate at the average value of the distribution. This reduces the problem to one similar to that of inferring $Z$ for a fixed $C$, as in Section 4.1. We therefore make the same assumption that $q(Z)$ factorizes according to Eq. (9). This gives updates of the form Eq. (16) where $\alpha$ has been set to its mean value.

$q(\alpha)$ **updates**

A fixed point condition for the optimum of Eq. (20) is

$$q(\alpha) \propto p(\alpha) e^{\langle \log p(A|Z,\alpha) \rangle_{q(Z)}},$$

Additionally we assume that $q(\alpha) = \prod_c q(\alpha_c)$. The resulting update

$$q(\alpha_c) \propto e^{\langle \log p(A|Z,\alpha) \rangle_{q(Z)} + \langle \log p(\alpha) \rangle_{\prod_{d \neq c} q(\alpha_d)}}$$

is difficult to compute and we take the naive approach of replacing averages by evaluation at the mean

$$q(\alpha_c) \propto p(\alpha_c, \langle \alpha_{\backslash c} \rangle) p(A| \langle z \rangle, \alpha_c, \langle \alpha_{\backslash c} \rangle) \quad (22)$$

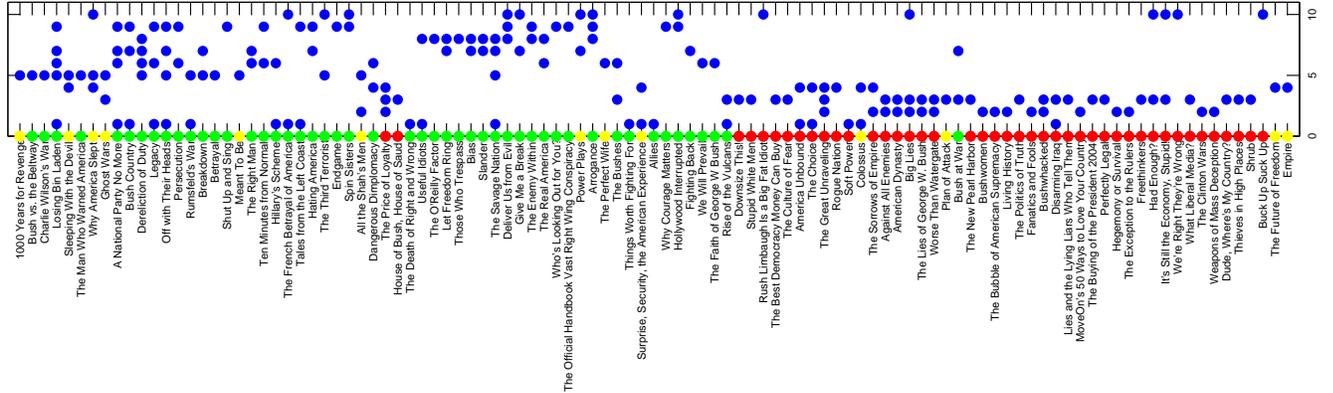

Figure 4: Political Books. Plotted is the $105 \times 10$ matrix $Z$, found by approximating the $105 \times 105$ adjacency co-bought matrix, where a dot indicates $q(z_{i,c}) > 0.5$. By inspection, cliques 5,6,7,8,9 largely correspond to 'conservative' books. Green indicates 'conservative' books, yellow 'neutral' and red 'liberal' books.

Since $\alpha_c$ is binary, we can easily find Eq. (22) by evaluating at its two states.

Formally, the prior $p(\alpha)$ requires $\alpha \in \{0,1\}^C$. However, in the above approximation, the mean $\alpha$ is non-binary. To deal with this, we replace $\sum_c I[\alpha_c = 1]$ by $\sum_c \langle \alpha_c \rangle$ and $\sum_c I[\alpha_c = 0]$ by $C_{max} - \sum_c \langle \alpha_c \rangle$. Since the expressions are valid for non-integer sums, this approximate procedure remains well defined.

The algorithm then updates $q(\alpha)$ and $q(Z)$ until convergence. The effect is that, beginning with $C_{max}$ clusters, under the updating, the posterior assigns $\alpha$'s not required to state zero.

## 6 Demonstrations

**DIMACS `MAX-CLIQUE`**

Our method aims to find a complete characterisation of an undirected graph into constituent clusters. By setting $\beta$ suitably high ($\beta = 10$ in the experiments), we impose that perfect cliques constitute clusters. In Fig.2a we show the adjacency matrix for a 200 vertex graph, taken from the DIMACS 1996 `MAX CLIQUE` challenge [4]. This graph was constructed to hide the largest clique in the graph and make it difficult to find based on the recursive algorithms of the time. Whilst more recent algorithms have been constructed that readily find the largest clique in this graph [13], this problem serves as an interesting baseline to see if our algorithm, in searching for a complete decomposition, also solves `MAX-CLIQUE` for this graph. Running our Mean-Field algorithm with $C_{max} = 2000$ results in a clique-decomposition, Fig.2b, containing 1102 cliques[5]. In Fig.2c we plot a log histogram of the cluster sizes, indicating that there is only a single largest clique of size 12. The largest clique in the graph is indeed 12[4].

**Political Books Clustering**

The data consists of 105 books on US politics sold by the online bookseller Amazon. Edges in graph $G$, Fig.3a, represent frequent co-purchasing of books by the same buyers, as indicated by the 'customers who bought this book also bought these other books' feature on Amazon[10]. Additionally, books are labelled 'liberal', 'neutral', or 'conservative' according to the judgement of a politically astute reader[6]. Running our algorithm with an initial $C_{max} = 200$ cliques, the posterior contains 142 cliques[7], Fig.3b, giving a perfect reconstruction of the adjacency $A$. For comparison, the incidence matrix has 441 2-cliques. To cluster the data more aggressively, we fix $C = 10$ and run our algorithm. As expected, this results only in an approximate clique decomposition, $A \approx H(ZZ^\mathsf{T})$, as plotted in Fig.3c. The resulting $105 \times 10$ approximate clique matrix is plotted in Fig. 4 and demonstrates how individual books are present in more than one cluster. Interestingly, the clusters found only on the basis of the adjacency matrix have some correspondence with the ascribed political leanings of each book.

## 7 Latent Parameterisations for Zero-Constrained Positive Matrices

We may use an undirected graph $G$ to represent zero constraints on a positive definite matrix $K$. In particular, missing edges in $G$ with adjacency $A_{ij} = 0$, correspond to zero entries $K_{ij} = 0$[8]. An example ap-plication would be to fit a Gaussian to data under the constraint that specified elements of the covariance are zero. In such cases, it is useful to have a parameterisation of the allowed space of covariances. We denote the space of positive definite matrices constrained through $G$ by $M^+(G)$. Our approach is based on the simple observation that by replacing non-zero entries of a clique matrix $Z$ with arbitrary real values, $Z \to Z^*$, the matrix $Z^*(Z^*)^{\mathsf{T}}$ is positive (semi) definite. An immediate question is the richness of such a parameterisation – can all of $M^+(G)$ be reached in this way?

---
[5]This takes roughly 30s using a 1Ghz machine.

[6]See www-personal.umich.edu/∼mejn/netdata/.

[7]This take roughly 10s on a 1GHz machine.

[8]In a Gaussian context, missing edges in $G$ typically correspond to missing edges in the inverse covariance. Much of our initial discussion relates only to constraining positive

### 7.1 Decomposable Case

For $G$ decomposable, parameterising $M^+(G)$ is straightforward[14, 12, 19]. For example one may appeal to the following:

**Theorem 1** (Paulsen et al., 1989). *The following are equivalent for an undirected graph $G$: (i) the graph is decomposable; (ii) there exists a permutation of the vertices such that with respect to this renumbering every $K \in M^+(G)$ factors as $K = T^{\mathsf{T}} T$ with $T \in M(G)$ and $T$ upper triangular.*

For decomposable $G$, provided the vertices are perfect elimination ordered, the Cholesky factor has the same structure as $G$[19]. In other words, provided the vertices are ordered correctly, the lower triangular part of the adjacency matrix is a clique matrix and furthermore parameterises all of $M^+(G)$. All positive definite matrices under decomposable zero-constraints can therefore be parameterised by some clique matrix.

**Definition 2** (Expanded Clique Matrix). Given a Clique Matrix $Z \in \{0,1\}^{V \times C}$, the Expanded Clique matrix consists of $Z$ appended with columns corresponding to all unique sub-columns of $Z$. A subcolumn of $z^c$ is defined by replacing one or more entries containing $z_i^c = 1$ by $z_i^c = 0$.

The expanded Clique Matrix corresponding to the minimal clique matrix derived from Fig.1b is

$$\begin{pmatrix} 1 & 0 \\ 1 & 1 \\ 1 & 1 \\ 0 & 1 \end{pmatrix} \to \begin{pmatrix} 1 & 0 & 1 & 1 & 0 & 0 & 0 & 1 & 0 & 0 & 0 \\ 1 & 1 & 1 & 0 & 1 & 1 & 0 & 0 & 1 & 0 & 0 \\ 1 & 1 & 0 & 1 & 1 & 0 & 1 & 0 & 0 & 1 & 0 \\ 0 & 1 & 0 & 0 & 0 & 1 & 1 & 0 & 0 & 0 & 1 \end{pmatrix}$$
(23)

In the above, the expansion is ordered such that all 3-cliques are enumerated, then all 2-cliques and finally all 1-cliques.

Starting from a minimal clique matrix for a decomposable graph, the expansion of this minimal clique matrix must contain all the columns of the Cholesky

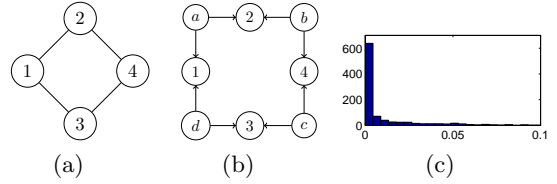

Figure 5: (a) Non-decomposable graph. (b) Correlations can be induced via latent variables.(c) Histogram of the rms errors in approximating covariances according to graph (a) with an expanded incidence matrix.

factor $T^{\mathsf{T}}$. For the example for $G$ in Fig.1b, the lower triangular Cholesky factor is[9]

$$\begin{pmatrix} * & 0 & 0 & 0 \\ * & * & 0 & 0 \\ * & * & * & 0 \\ 0 & * & * & * \end{pmatrix}$$

which corresponds to columns $1, 2, 7, 11$ of the expanded clique matrix, Eq. (23). Clearly, in general, the expanded clique matrix is an over-parameterisation of $M^+(G)$ for decomposable $G$.

### 7.2 Non-decomposable Case

For $G$ non-decomposable, no explicit parameterisation is generally possible and techniques based on Positive Definite matrix completion are required[12, 18, 5, 14]. For the specific example in Fig.5a, the lower Cholesky factor has the form

$$\begin{pmatrix} c_{11} & 0 & 0 & 0 \\ c_{21} & c_{22} & 0 & 0 \\ c_{31} & c_{32} & c_{33} & 0 \\ 0 & c_{42} & c_{43} & c_{44} \end{pmatrix}, \text{ with } c_{21}c_{31} + c_{22}c_{32} = 0$$
(24)

which can be found explicitly in this case. However, more generally, for non-decomposable graphs, one cannot identify those elements of the Cholesky factor which may be set to zero, with the remainder determined by the positivity requirement[14, 19].

An alternative is to use latent variables to explicitly parameterise $M^+(G)$. One may use Factor Analysis[1]

$$x = F\epsilon, \ \epsilon \sim \mathcal{N}(0, I) \quad \Rightarrow \Sigma = FF^{\mathsf{T}}$$

where the factor matrix $F$ is suitably structured in order to force zeros in specific elements of $\Sigma^{10}$.

---

[9]In general, for a matrix with elements $d_{ij} \in \{0, 1\}$, we use $D^*$ to denote a matrix with $d_{ij}^* = 0$ if $d_{ij} = 0$, and arbitrary values elsewhere.

[10]By writing $F = [\tilde{F}|D]$ where $D$ is diagonal, this is explicitly Factor Analysis. Unlike standard FA, the matrix $\tilde{F}$ will typically be non-square and sparse.

definite matrices and is independent of its application.

A special case of the above is to use a latent variable to induce correlation between $x_1$ and $x_2$ via a local Directed Graph element $x_1 \leftarrow \epsilon_{12} \rightarrow x_2$. For each edge in $G$, a corresponding latent $\epsilon$ can thus be introduced to form correlations between all pairs of variables, without introducing correlations on missing edges in $G$[8]. By taking $F = [Z^*_{inc}|I^*]$, it is clear that this latent variable approach (see, for example, [16]) is reproduced and is a special case of restricting Cliques to Incidence matrices.

To show that not all of $M^+(G)$ can be reached by clique matrices, consider Fig.5a. In this particularly simple case, the minimal clique matrix is the same as the incidence matrix, and the expanded clique matrix is simply the incidence matrix with the identity matrix appended. In this case, therefore, the expanded clique matrix contains columns with only two non-zero entries. However, the Cholesky factor Eq. (24) contains columns with 3 non-zero entries, so that there is no immediate assignment of $[Z^*_{inc}|I^*]$ which will match the Cholesky factor.

For the non-decomposable graph 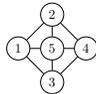 the minimal clique matrix contains 3-cliques so that its expansion contains columns that an expansion based on an incidence matrix would not. In this case our approximate parameterisation is therefore richer than would be obtained from simply introducing a latent auxiliary variable for each edge of the graph[8, 15].

### 7.3 Maximum Likelihood Solution

In fitting a Gaussian $\mathcal{N}(0, \Sigma)$ to zero mean data, with sample covariance $S$, the ML solution minimises

$$\kappa(\Sigma) \equiv \text{Tr}\left(\Sigma^{-1} S\right) + \log \det \Sigma \qquad (25)$$

Our interest is to minimize $\kappa$ subject to zero constraints on $\Sigma$ specified through $G$, with $\sigma_{ij} = 0$ if $A_{ij} = 0$. For $G$ decomposable, the problem is essentially trivial, since $M^+(G)$ is easily characterized via a structured Cholesky factor, $\Sigma \equiv C^\mathsf{T}(\theta)C(\theta)$ see for example [14], for which one can parameterise Eq. (25) using $\kappa(\theta)$ and perform unconstrained minimisation over the free parameters $\theta$ of the Cholesky factor.

In the non-decomposable $G$ case, no explicit parameterisation of $M^+(G)$ is feasible. A common approach in this case is to recognise that solutions to this satisfy $\left[\Sigma^{-1}_{ij}\right] = \left[\Sigma^{-1} S \Sigma^{-1}\right]_{ij}$ for $A_{ij} = 1$ and $\sigma_{ij} = 0$ otherwise[2] and define iterative procedures to solve this equation[7]. Alternatively, Positive Definite Completion methods may be used to parameterise $M^+(G)$. Our approach uses the parameterisation $\Sigma = Z^*(Z^*)^\mathsf{T}$ where $Z$ should be chosen as large as can be computationally afforded. $Z$ can be determined by running the algorithm of Section 4.1[11]. Although for non-decomposable $G$, not all of $M^+(G)$ is guaranteed reachable through this parameterisation, one may expect that numerically this may be sufficiently close. A benefit of this approach is that one may then minimize Eq. (25) with respect to the free parameters of $Z^*$ using any standard optimisation technique, and convergence is guaranteed. Since our parameterisation has a natural latent variable representation (it is a form of structured Factor Analysis), EM and Bayesian techniques can also be used in this case. A numerical example is plotted in Fig.5c. We take the $4 \times 8$ expanded clique matrix corresponding to Fig.5a and minimise Eq. (25) with respect to the non-zero entries of the clique matrix[12]. Each sample matrix $S$ is generated randomly by drawing values of the Cholesky factor Eq. (24) independently from a zero mean unit variance Gaussian. In Fig.5c we plot the root mean square error between the learned $\Sigma$ and sample covariance $S$, averaged over all non-zero components of $\Sigma$. The histogram of the error, computed from 1000 simulations shows that, whilst a few have appreciable error, the vast majority of cases are numerically well approximated by the expanded clique matrix technique, even though the graph $G$ is non-decomposable.

## 8 Summary

We introduced a graph matrix decomposition based on an extension of the incidence matrix concept. Finding the clique decomposition corresponding to the smallest number of cliques is a hard problem, and we considered a relaxed version of the problem to find an approximate clique decomposition based on a variational algorithm. The approach can be seen as a form of binary factorisation of the adjacency matrix[13]. Clear extensions of this work would be to consider alternative approximate inference schemes, including sampling methods, for which 'infinite' extensions are also available[11]. An application of clique matrices is to parameterising positive definite matrices under specified zero constraints. We showed that constraints corresponding to decomposable graphs trivially admit a clique matrix representation, and how our structured Factor Analysis technique can be used to approximate the non-decomposable case. This is a richer parameterisation than those latent models which consider only pairwise correlations in forming the latent model. Indeed, the so-called ancillary variable technique is a special case of using incidence, as opposed to clique matrices. The latent variable formulation additionally offers an alternative to recent works on conjugate priors for constrained covariances in Bayesian learning.

---

[11]A heuristic is to initialise $Z$ for the variational algorithm based on the lower Cholesky factor of a spanning decomposable graph, augmented with missing two cliques. Once the algorithm converges to an approximate minimal clique matrix, its expansion is used to form the parameterisation $Z^*$.

[12]We chose this simple case since the exact parameterisation of all $M^+(G)$ is easy to write down. Whilst here the expanded clique and incidence matrices are equivalent, the reader should bear in mind that in more complex situations, the expansion based on a clique matrix provides a richer parameterisation than that of the incidence matrix.

[13]A parallel development to our own work is [11], which considers binary factorisation of more general matrices. Thanks to Zoubin Ghahramani for pointing me to this.

C-code for clique matrices is available from the author.

**Acknowledgements**

I thank R. Silva, T. Cemgil and M. Herbster for discussions. Supported in part by the European PASCAL Network of Excellence, IST-2002-506778.